\documentclass[doublecol,amsmath,amssymb]{epl2}
\usepackage{amssymb}
\usepackage{graphicx}
\usepackage{dcolumn}
\usepackage{bm}
\usepackage{hhline}
\usepackage{longtable}

\begin{document}

\title{Maximal planar scale-free Sierpinski networks with small-world effect and power-law strength-degree correlation}
\shorttitle{Maximal planar scale-free Sierpinski networks with
small-world effect and power-law strength-degree correlation}

\author{Zhongzhi Zhang\inst{1,2} \footnote{ \email{zhangzz@fudan.edu.cn (Z.Z. Zhang)} }\and Shuigeng Zhou\inst{1,2} \footnote{\email{sgzhou@fudan.edu.cn (S.G. Zhou)}} \and Lujun Fang\inst{1,2} \and Jihong Guan\inst{3} \and Yichao Zhang\inst{4}}
\shortauthor{Zhongzhi Zhang, Shuigeng Zhou, Lujun Fang, Jihong Guan,
Yichao Zhang}

 \institute{
  \inst{1} Department of Computer Science and Engineering,
Fudan University, Shanghai 200433, China\\
  \inst{2} Shanghai Key Lab of Intelligent Information
Processing, Fudan University, Shanghai 200433, China\\
 \inst{3} Department of Computer Science and Technology,
Tongji University, 4800 Cao'an Road, Shanghai 201804, China\\
 \inst{4} Material and Engineering Institute, Shanghai
University, Shanghai 200072, China}

\date{\today}

\begin{abstract}{
Many real networks share three generic properties: they are
scale-free, display a small-world effect, and show a power-law
strength-degree correlation. In this paper, we propose a type of
deterministically growing networks called Sierpinski networks, which
are induced by the famous Sierpinski fractals and constructed in a
simple iterative way. We derive analytical expressions for degree
distribution, strength distribution, clustering coefficient, and
strength-degree correlation, which agree well with the
characterizations of various real-life networks. Moreover, we show
that the introduced Sierpinski networks are maximal planar graphs.}
\end{abstract}



 \maketitle

\section{Introduction}

In the last few years, research of complex networks have become a
focus of attention from the scientific
community~\cite{AlBa02,DoMe02,Ne03,BoLaMoChHw06,CoRoTrVi07}. One of
the main reasons behind the popularity of  complex networks is their
flexibility and generality for representing real systems in nature
and society. Researchers have done a lot of empirical studies,
uncovering that various real-life networks sharing some generic
properties: power-law degree distribution~\cite{BaAl99}, small-world
effect including small average path length (APL) and high clustering
coefficient~\cite{WaSt98}. Recently, many authors have described
some real-world systems in terms of weighted networks, where an
interesting empirical phenomenon has been observed that there exists
a power-law scaling relation between the strength $s$ and degree $k$
of nodes, i.e. $s\sim k^{\beta}$ with
$\beta>1$~\cite{BaBaPaVe04,LiCa04,GabaCaSeCa05,SeBoDi05}.

With the intention of studying the above properties of real-world
systems, a wide variety of models have been
proposed~\cite{AlBa02,DoMe02,Ne03,BoLaMoChHw06}. Watts and Strogatz,
in their pioneering paper, introduced the famous small-world network
model (WS model)~\cite{WaSt98}, which exhibits small APL and high
clustering coefficient. Another well-known model is Barab\'asi and
Albert's scale-free network model (BA model)~\cite{BaAl99}, which
has a degree distribution of power-law form. However, in these two
elegant models, scale-free feature and high clustering are
exclusive. Driven by the two seminal papers~\cite{WaSt98,BaAl99}, a
considerable number of other models have been developed that may
represent processes more realistically taking place in real-world
networks~\cite{DoMeSa00,AlBa00,DoMe00b,BiBa01,HoKi02,KlEg02a,XuLiWu06}.
Very recently, Barrat, Barth\'elemy, and Vespignani have introduced
a model (BBV) for the growth of weighted
networks~\cite{BaBaVe04a,BaBaVe04b}, which is the first weighted
network model that yields a scale-free behavior for strength and
degree distributions. Enlightened by BBV's remarkable work, various
weighted network models have been proposed to explain the properties
found in real
systems~\cite{Bi05,WaWaHuYaQu05,WaHuZhWaXi05,GoKaKi05,MuMa06,LiWaFaDiWu06,QuJiZhWaYi07}.
These models may give some insight into the realities. Particulary,
some of them presente all the above-mentioned three characteristics
such as power-law degree distribution, small-world effect, and
power-law strength-degree
correlation~\cite{Bi05,WaWaHuYaQu05,WaHuZhWaXi05}. Although great
progresses have been made in the research of network topology,
modeling complex networks with general structural properties is
still of current interest.

On the other hand, fractals are an important tool for the
investigation of physical phenomena~\cite{Ma82}. They were used to
describe physical characteristics of things in nature and life
systems such as clouds, trees, mountains, rivers, coastlines, waves
on a lake, bronchi, and the human circulatory system, to mention but
a few. A vast literature on the theory and application of fractals
has appeared. Among many deterministic and statistical fractals, the
Sierpinski gasket~\cite{Si15,Re94} is one of the earliest
deterministic fractals; it has provided a rich source for examples
of fractal behavior~\cite{Ma82,BrRe95,Ne06}. Our initial physical
motivation for this work lies in the use of the Sierpinski gasket as
models for complex networks.

In this letter, based on the well-known Sierpinski family fractals,
we introduce a class of deterministic networks, named Sierpinski
networks. We propose a minimal iterative algorithm for constructing
the networks and studying their structural properties. The networks
are maximal planar graphs, show scale-free distributions of degree
and strength, exhibit small-world effect, and display power-law
strength-degree correlation, which may provide valuable insights
into the real-life systems.

\section{The network derived from Sierpinski gasket}
We first introduce a family of fractals, called Sierpinski fractals,
by generalizing the construction of the Sierpinski gasket. The
classic Sierpinski gasket, shown in Fig 1(a), is constructed as
follows~\cite{Si15}. We start with an equilateral triangle, and we
denote this initial configuration by generation $t=0$. Then in the
first generation $t=1$, the three sides of the equilateral triangle
are bisected and the central triangle removed. This forms three
copies of the original triangle, and the procedure is repeated
indefinitely for all the new copies. In the limit of infinite $t$
generations, we obtain the well-known Sierpinski gasket denoted by
$SG_2(t)$. Another fractal based on the equilateral triangle can be
obtained if we perform a trisection of the sides and remove the
three down pointing triangles, to form six copies of the original
triangle. Continue this procedure in each copy recursively to form a
gasket, denoted $SG_3(t)$, shown in Fig. 1(b). Indeed, this can be
generalized to $SG_\omega(t)$, for any positive integer $\omega$
with $\omega \neq 1$, by dividing the sides in $\omega$, joining
these points and removing all the downward pointing
triangles~\cite{Ha92}. Thus, we obtain a family of fractals
(Sierpinski fractals), whose Hausdorff dimension is
$d_{f}=\log[\frac{1}{2}\omega(\omega+1)]/\log(\omega)$, which tends
to 2 as $\omega\rightarrow \infty$~\cite{Hu81}.

\begin{figure}
\begin{center}
\includegraphics[width=0.45\textwidth]{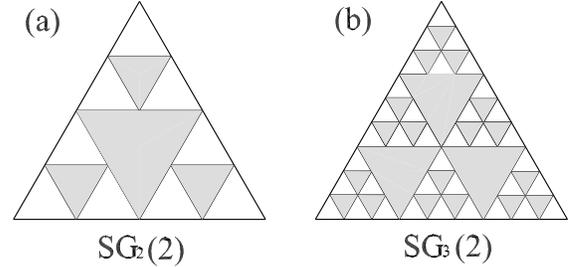} \
\end{center}
\caption[kurzform]{\label{Fig1} The first two stages in the
construction of two fractals from the family of Sierpinski gaskets.
}
\end{figure}

\begin{figure}
\begin{center}
\includegraphics[width=0.45\textwidth]{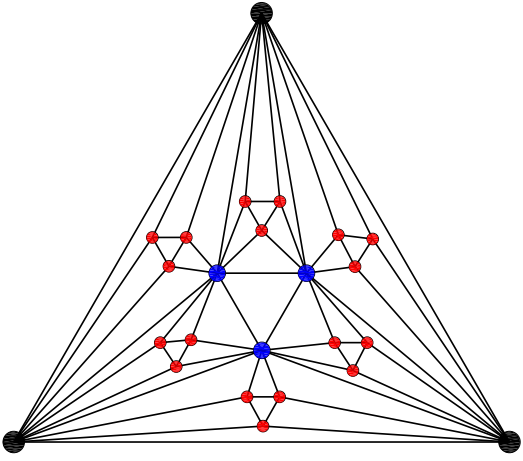} \
\end{center}
\caption[kurzform]{\label{network} Illustration of a deterministic
network in relation to $SG_3(2)$. }
\end{figure}

One can use these fractals to construct networks. The translation
from the fractals to network generation is quite straightforward.
Let the nodes (vertices) of the networks correspond to the removed
triangles and make two nodes connected if the boundaries of the
corresponding triangles contact one another. For uniformity, the
three sides of the initial equilateral triangle at step 0 also
correspond to three different nodes. Figure 2 shows a network based
on $SG_3(2)$. All generated networks have similar properties. In
what follows we will focus on the particular network corresponding
to the case of $\omega=3$ (see Figure~\ref{network}), which is also
called Sierpinski network, and other cases can be treated
analogously.

\section{Iterative algorithm for the maximal planar network}

In the construction process of $SG_3(t)$, for each equilateral
triangle at arbitrary generation, once we perform a trisection of
its sides and remove the three resultant down pointing triangles,
six copies of it are formed. When building the network, it is
equivalent that for every group of three new added nodes, six new
small equilateral triangles are generated, each of which may create
three
 nodes in the next generation. According to this, we can
introduce a general algorithm to create the corresponding network,
denoted by $F(t)$ after $t$ generation evolutions.

\begin{figure}
\begin{center}
\includegraphics[width=0.4\textwidth]{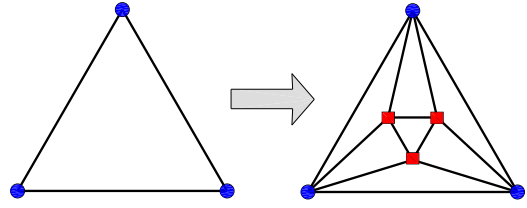}
\end{center}
\caption[kurzform]{\label{iterative} Iterative construction method
for the network. }
\end{figure}

The iterative algorithm for the network is as follows: For $t=0$,
$F(0)$ consists of three nodes forming a triangle. Then, we add
three nodes into the original triangle. These three new nodes are
linked to each other shaping a new triangle, and both ends of each
edge of the new triangle are connected to a node of the original
triangle. Thus we get $F(1)$, see Figure~\ref{iterative}. For $t\geq
1$, $F(t)$ is obtained from $F(t-1)$. For each of the existing
triangles of $F(t-1)$ that is not composed of three simultaneously
emerging nodes and has never generated a node before, we call it an
active triangle. We replace each of the existing active triangles of
$F(t-1)$ by the connected cluster on the right hand of Figure 3 to
obtain $F(t)$. The growing process is repeated until the network
reaches a desired order (node number of network).
Figure~\ref{network} shows the network growing process for the first
two steps.

Next we compute the order and size (number of all edges) of the
network $F(t)$. Let $L_v(t)$, $L_e(t)$ and $L_\Delta(t)$ be the
number of vertices, edges and active triangles created at step $t$,
respectively. By construction (see also Figure~\ref{iterative}),
each active triangle in $F(t-1)$ will be replaced by six active
triangles in $F(t)$. Thus, it is not difficult to find the following
relation: $L_\Delta(t)=6\,L_\Delta(t-1)$. Since $L_\Delta(0)=1$, we
have $L_\Delta(t)=6^{t}$.

Note that each active triangle in $F(t-1)$ will lead to an addition
of three new nodes and nine new edges at step $t$, then one can
easily obtain the following relations:
$L_v(t)=3\,L_\Delta(t-1)=3\cdot6^{t-1}$, and
$L_e(t)=9\,L_\Delta(t-1)=9\cdot6^{t-1}$ for arbitrary $t>0$. From
these results, we can compute the order and size of the network. The
total number of vertices $N_t$ and edges $E_t$ present at step $t$
is
\begin{equation}\label{Nt}
N_t=\sum_{t_i=0}^{t}L_v(t_i)=\frac{3\cdot6^{t}+12}{5}
\end{equation}
and
\begin{equation}\label{Et}
E_t =\sum_{t_i=0}^{t}L_e(t_i)=\frac{9\cdot6^{t}+6}{5},
\end{equation}
respectively. So for large $t$, the average degree $\overline{k}_t=
\frac{2E_t}{N_t}$ is approximately $6$, which shows the network is
sparse as most real systems.

From Eqs.~(\ref{Nt}) and~(\ref{Et}), we have $E_t=3N_{t}-6$. In
addition, by the very construction of the network, it is obvious
that arbitrary two edges in the network never cross each other. Thus
our network is a maximal planar network (or graph)~\cite{We01},
which is similar to some previously studied
networks~\cite{AnHeAnSi05,ZhCoFeRo06,ZhYaWa05,ZhZhRo06}.

\section{Relevant characteristics of the network}
Now we study the statistical properties of the network, in terms of
degree distribution, clustering coefficient, average path length,
and strength distribution.

\subsection{Degree distribution}

When a new node $i$ is added to the graph at step $t_i$ ($t_i\geq
1$), it has a degree of $4$. Let $L_\Delta(i,t)$ be the number of
active triangles at step $t$ that will create new nodes connected to
the node $i$ at step $t+1$. Then at step $t_i$, $L_\Delta(i,
t_i)=3$. From the iterative generation process of the network, one
can see that at any step each two new neighbors of $i$ generate
three new active triangles with involving $i$, and one of its
existing active triangle is deactivated simultaneously. We define
$k_i(t)$ as the degree of node $i$ at time $t$, then the relation
between $k_i(t)$ and $L_\Delta(i,t)$ satisfies:
\begin{equation}\label{deltak}
L_\Delta(i,t)=k_i(t)-1.
\end{equation}
Now we compute $L_\Delta(i,t)$. By construction,
$L_\Delta(i,t)=3\,L_\Delta(i,t-1)$. Considering the initial
condition $L_\Delta(i, t_i)=3$, we can derive
$L_\Delta(i,t)=3^{t-t_{i}+1}$. Then at time $t$, the degree of
vertex $i$ becomes
\begin{equation}\label{ki}
k_i(t)=3^{t-t_{i}+1}+1.
\end{equation}

It should be mentioned that the initial three vertices created at
step 0 have a little different evolution process from other ones. We
can easily obtain: $L_\Delta(0,t)=3^{t}$ and $k_i(t)=3^{t}+1$.  So
at step $t$, initial three vertices have the same degrees as those
ones born at step 1.

Equation~(\ref{ki}) shows that the degree spectrum of the network is
discrete.  It follows that the cumulative degree
distribution~\cite{Ne03} is given by
\begin{equation}\label{pcumk}
P_{\rm cum}(k)=\sum_{\tau \leq t_i}\frac{L_v(\tau)}{N_t}
={3\cdot6^{t_i}+12\over 3\cdot6^{t}+12}.
\end{equation}
Substituting for $t_i$ in this expression using
$t_i=t+1-\frac{\ln(k-1)}{\ln 3}$ gives
\begin{equation}
P_{\rm
cum}(k)=\frac{18\cdot6^{t}\cdot(k-1)^{-(\ln6/\ln3)}+12}{3\cdot6^{t}+12}.
\end{equation}
When $t$ is large enough, one can obtain
\begin{equation}\label{gammak}
P_{\rm cum}(k)=6\cdot(k-1)^{-\left[1+(\ln2/\ln3)\right]}.
\end{equation}
So the degree distribution follows a power law form with the
exponent $\gamma_{k}=2+\frac{\ln2}{\ln3}$.

\subsection{Clustering coefficient}

The clustering coefficient~\cite{WaSt98} $ C_i $ of node $i$ is
defined as the ratio between the number of edges $e_i$ that actually
exist among the $k_i $ neighbors of node $i$ and its maximum
possible value, $ k_i( k_i -1)/2 $, i.e., $ C_i =2e_i/[k_i( k_i
-1)]$. The clustering coefficient of the whole network is the
average of $C_i^{'}s $ over all nodes in the network.

For our network, the analytical expression of clustering coefficient
$C(k)$ for a single node with degree $k$ can be derived exactly.
When a node enters the system, both $k_i$ and $e_i$ are 4. In the
following iterations, each of its active triangles increases both
$k_{i}$ and $e_{i}$ by 2 and 3, respectively. Thus, $e_{i}$ equals
to $4+\frac{3}{2}\left(k_{i}-4\right)$ for all nodes at all steps.
So one can see that there exists a one-to-one correspondence between
the degree of a node and its clustering. For a node of degree $k$,
we have
\begin{equation}\label{Ck}
C(k)= \frac{2\,e}{k(k-1)}=
\frac{2\left[4+\frac{3}{2}(k-4)\right]}{k(k-1)}=\frac{4}{k}-\frac{1}{k-1}.
\end{equation}
In the limit of large $k$, $C(k)$ is inversely proportional to
degree $k$. The same scaling of $C(k)\sim k^{-1}$ has also been
observed in several real-life networks~\cite{RaBa03}.

\begin{figure}
\begin{center}
\includegraphics[width=0.45\textwidth]{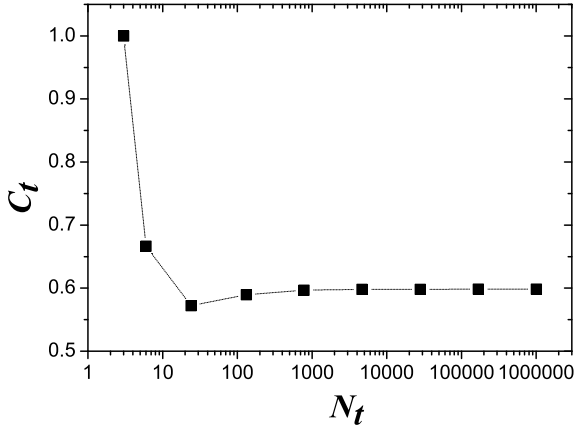} \
\end{center}
\caption[kurzform]{\label{clustering} Semilogarithmic plot of
average clustering coefficient $C_t$ versus network order $N_{t}$.}
\end{figure}

Using Eq. (\ref{Ck}), we can obtain the clustering $C_t$ of the
networks at step $t$:
\begin{equation}\label{ACC}
C_t=
    \sum_{r=0}^{t}\left[\frac{
    L_v(r)}{N_{t}}\left(\frac{4}{D_r}-\frac{1}{D_r-1}\right)\right],
\end{equation}
where the sum runs over all the nodes
and $D_r$ 
is the degree of the nodes created at step $r$, which is given by
Eq.~(\ref{ki}). In the infinite network order limit
($N_{t}\rightarrow \infty$), Eq.~(\ref{ACC}) converges to a nonzero
value $C=0.598$, as shown in Fig.~\ref{clustering}. Therefore, the
average clustering coefficient of the network is very high.

\subsection{Average path length}

Shortest paths play an important role both in the transport and
communication within a network and in the characterization of the
internal structure of the network.  We represent all the shortest
path lengths of $F(t)$ as a matrix in which the entry $d_{ij}$ is
the geodesic path from node $i$ to node $j$, where geodesic path is
one of the paths connecting two nodes with minimum length. The
maximum value of $d_{ij}$ is called the diameter of the network. A
measure of the typical separation between two nodes in $F(t)$ is
given by the average path length $d_{t}$, also known as
characteristic path length, defined as the mean of geodesic lengths
over all couples of nodes.

\begin{figure}
\begin{center}
\includegraphics[width=0.45\textwidth]{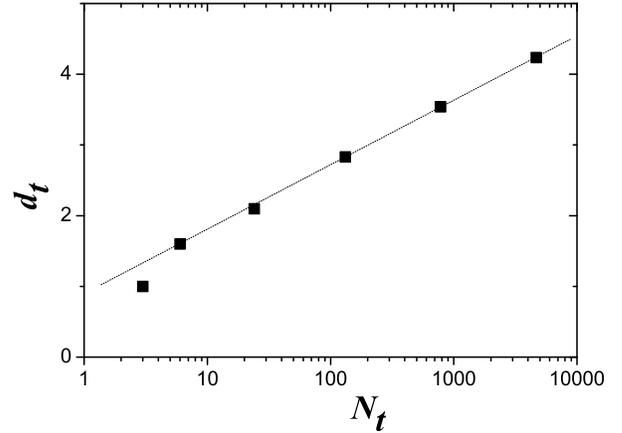} \
\end{center}
\caption[kurzform]{\label{distance} Average path length $d_{t}$
versus network order $N_{t}$ on a semilogarithmic scale. The solid
line is a guide to the eye.}
\end{figure}

In Fig.~\ref{distance}, we report the dependence relation of APL
$d_{t}$ on network size $N_{t}$. From Fig.~\ref{distance}, one can
see that the average path length grows logarithmically with
increasing size of the network. This logarithmic scaling of $d_{t}$
with network size $N_{t}$, together with the large clustering
coefficient obtained in the preceding subsection, shows that the
considered graph has a small-world effect.

\subsection{Strength distribution and strength-degree correlation}

Strength usually represents resources or substances allocated to
each node, such as wealth of individuals in financial contact
networks~\cite{XiWaHuZh05}, the number of passengers in airports in
world-wide airport networks~\cite{GuMoTuAm05}, the throughput of
power stations in electric power grids~\cite{AlAlNa04}, and so on.
In our model, the strength of a node is defined as the area of the
removed triangle it corresponds to~\cite{Sc01}. For uniformity, let
the initial three nodes born at step 0 have the same strength as
those created at step 1.

We assume that the area of the initial equilateral triangle of the
Sierpinski gasket is $\mathbb{A}$. By the very construction of the
network, all simultaneously emerging nodes have the same strength,
because their corresponding triangles have identical area. It is
easy to find that each removed triangle covers the portion
$\mu=\frac{1}{9}$ of one removed triangle in the preceding
generation. After $t$ iterations, all nodes which are generated at a
certain step $t_i = 1,2, \cdots ,t$ have the strength $s_i(t)$:
\begin{equation}\label{s01}
s_i(t)=\mu^{t_i}\cdot\mathbb{A},
\end{equation}
from which we have
\begin{equation}\label{s02}
s_i(t)=\mu^{t_i-t}\cdot \bigtriangleup,
\end{equation}
where $\bigtriangleup$ is the area $s_t(t)$ of a triangle removed at
step $t$.

Eqs.~(\ref{ki}) and~(\ref{s02}) yield a power-law correlation
between strength and degree of a node:
\begin{equation}\label{s03}
s_i(t)=\mu\bigtriangleup\cdot
(k_i-1)^{\ln(1/\mu)/\ln3}=\frac{\bigtriangleup}{9} (k_i-1)^2,
\end{equation}
which implies when $k$ is large enough, $s\sim k^2$. This nontrivial
power-law scaling between strength of a node and its degree has been
empirically observed in a variety of real networks, such as the
airport networks, the shareholder networks, and the Internet.

Analogously to computation of degree distribution, we one can find
that the strength distribution is also scale-free with exponent
$\gamma_{s}$ as:
\begin{equation}\label{gammas01}
\gamma_{s}=\frac{3}{2}+\frac{\ln2}{2\ln3}.
\end{equation}

As known to us all, for weighted networks with non-linear
strength-degree correlation $s\sim k^\beta$, if their distributions
of degree and strength behave as power laws,  $P(k)\sim
k^{-\gamma_{k}}$ and $P(s)\sim s^{-\gamma_{s}}$, then there is a
general relation between $\gamma_{k}$ and $\gamma_{s}$
as~\cite{BaBaVe04a,BaBaVe04b}:
\begin{equation}\label{gammas02}
\gamma_{s}=\frac{\gamma_{k}}{\beta}+\frac{\beta-1}{\beta}.
\end{equation}
We have shown that in our model $\gamma_{k}=2+\frac{\ln2}{\ln3}$ and
$\beta=2$. According to Eq.~(\ref{gammas02}), the exponent of
strength distribution is
$\gamma_{s}=\frac{3}{2}+\frac{\ln2}{2\ln3}$, giving the same
$\gamma_{s}$  value as that obtained in the direct calculation of
the strength distribution, see equation (\ref{gammas01}).

\section{Conclusion}

Deterministic model makes it easier to gain a visual understanding
of how do different nodes relate to each other forming complex
networks~\cite{BaRaVi01,DoGoMe02,CoFeRa04,ZhRoZh07,
HiBe06,ZhZhZo07}. On the basis of Sierpinski fractals, we have
proposed and studied a kind of deterministic networks. According to
the network construction processes we have presented an algorithm to
generate the networks, based on which we have obtained the
analytical results for degree distribution, clustering coefficient,
strength distribution, as well as strength-degree correlation. We
have shown that the networks have three important properties:
power-law distributions, small-world effect, and power-law
strength-degree correlation, which are in good accordance with a
variety of real-life networks. In addition, the networks are maximal
planar graphs, which may be helpful for designing printed circuits.

Although we have studied only a particular network, in a similar
way, one can easily investigate other Sierpinski networks with
various values of $\gamma_s$ and $\gamma_k$, and their general
properties such as small-world effect and power-law strength-degree
relation are similar. Moreover, using the idea presented, one can
also establish random networks, which display similar features as
their deterministic counterparts studied here. As the classic
Sierpinski gaskets are important for the understanding of
geometrical fractals in real systems, we believe that our research
could be useful in the understanding and modeling of real-world
networks.

\section{Acknowledgment}
This research was supported by the National Natural Science
Foundation of China under Grant Nos. 60496327, 60573183, and
90612007, the Postdoctoral Science Foundation of China under Grant
No. 20060400162, and the Program for New Century Excellent Talents
in University of China (NCET-06-0376). Z.Z. Zhang also acknowledges
the support from the Huawei Foundation of Science and Technology.


\begin{thebibliography}{ref1}
\bibitem{AlBa02} R. Albert and A.-L. Barab\'asi,
       Rev. Mod. Phys. {\bf 74}, 47 (2002).

\bibitem{DoMe02} S. N. Dorogvtsev and J.F.F. Mendes,
Adv. Phys. {\bf 51}, 1079 (2002).

\bibitem{Ne03} M. E. J. Newman,
SIAM Review {\bf 45}, 167 (2003).


\bibitem{BoLaMoChHw06}
S. Boccaletti, V. Latora, Y. Moreno, M. Chavezf, and D.-U. Hwanga,
Phy. Rep. {\bf 424}, 175 (2006).

\bibitem{CoRoTrVi07}
L. da. F. Costa, F. A. Rodrigues, G. Travieso, and P. R. V. Boas,
Adv. Phys. {\bf 56}, 167 (2007).


\bibitem{BaAl99} A.-L. Barab\'asi and R. Albert,
       Science {\bf 286}, 509 (1999).

\bibitem{WaSt98} D.J. Watts and H. Strogatz,
        Nature (London) {\bf 393}, 440 (1998).

\bibitem{BaBaPaVe04}
A. Barrat, M. Barth\'elemy, R. Pastor-Satorras, and A. Vespignani,
Proc. Natl. Acad. Sci. U.S.A. {\bf 101}, 3747 (2004).

\bibitem{LiCa04}
W. Li, and X. Cai, Phys. Rev. E {\bf 69}, 046106 (2004).

\bibitem{GabaCaSeCa05}
 D. Garlaschelli, S. Battiston, M.o Castri, V. D. P. Servedio, and G.
 Caldarelli,
 Physica A  {\bf 350}, 491 (2005).


\bibitem{SeBoDi05}
M. \'A. Serrano, M. Bogu\~{n}\'a, and A. D\'{\i}az-Guilera, Phys.
Rev. Lett. {\bf 94}, 038701 (2005).

\bibitem{DoMeSa00}
S. N. Dorogovtsev, J. F. F. Mendes, and A. N. Samukhin,
Phys. Rev. Lett. {\bf 85}, 4633 (2000).


\bibitem{AlBa00}
R. Albert, and A.-L. Barab\'asi,
Phys. Rev. Lett. {\bf 85}, 5234 (2000).

\bibitem{DoMe00b}
S. N. Dorogovtsev, and J. F. F. Mendes,
Europhys. Lett. {\bf 52}, 33 (2000). 

\bibitem{BiBa01}
G. Bianconi, and A.-L. Barab\'asi,
Europhys. Lett. {\bf 54}, 436 (2001). 

\bibitem{HoKi02}
P. Holme and B. J. Kim, Phys. Rev. E {\bf 65}, 026107 (2002).

\bibitem{KlEg02a}
K. Klemm and V. M. Egu\'iluz, Phys. Rev. E {\bf 65}, 036123 (2002).

\bibitem{XuLiWu06}
Q. Xuan, Y. Li, and T.-J. Wu, Phys. Rev. E {\bf 73}, 036105 (2006).

\bibitem{BaBaVe04a}
A. Barrat, M. Barth\'elemy, and A. Vespignani, Phys.
Rev. Lett. {\bf 92}, 228701 (2004).

\bibitem{BaBaVe04b}
A. Barrat, M. Barth\'elemy, and A. Vespignani, Phys.
Rev. E {\bf 70}, 066149 (2004).

\bibitem{Bi05}
G. Bianconi, Europhys. Lett. {\bf 71}, 1029 (2005). 

\bibitem{WaWaHuYaQu05}
W. X. Wang, B. H. Wang, B. Hu, G. Yan, and Q. Ou, Phys. Rev. Lett.
\textbf{94}, 188702 (2005).


\bibitem{WaHuZhWaXi05}
W. X. Wang, B. Hu, T. Zhou, B. H. Wang, and Y. B. Xie,
Phys. Rev. E \textbf{72}, 046140 (2005).


\bibitem{GoKaKi05}
K.-I. Goh, B. Kahng, and D. Kim, Phys. Rev. E {\bf 72}, 017103
(2005).

\bibitem{MuMa06}
G. Mukherjee and S. S. Manna, Phys. Rev. E {\bf 74}, 036111 (2006).

\bibitem{LiWaFaDiWu06}
M. Li, D. Wang, Y. Fang, Z. Di, and J. Wu, New J. Phys. \textbf{8},
72 (2006).

\bibitem{QuJiZhWaYi07}
Q. Ou, Y. D. Jin, T. Zhou, B. H. Wang, and B. Q. Yin, Phys. Rev. E
{\bf 75}, 021102 (2007).

\bibitem{Ma82}
B. B. Mandelbrot, {\em The Fractal Geometry of Nature} (Freeman, San
Francisco, CA, 1982).

\bibitem{Si15}
 W. Sierpinski, Comptes Rendus (Paris) {\bf 160}, 302
(1915).

\bibitem{Re94}
C. A. Reiter, Comput. $\&$ Graphics {\bf 18}, 885 (1994)

\bibitem{BrRe95}
G. F. Brisson, C. A. Reiter, Chaos, Solitons $\&$ Fractals {\bf 5},
2191 (1995).

\bibitem{Ne06}
G. R. Newkome, et al. Science {\bf 312}, 1782 (2006).


\bibitem{Ha92}
B.M. Hambly,
Probab. Theory Related Fields {\bf 94}, 1 (1992).

\bibitem{Hu81}
S. Hutchinson,
Indiana Univ. Math. J. {\bf 30}, 713 (1981).

\bibitem{We01} D.B. West,
    {\em Introduction to Graph Theory}
    (Prentice-Hall,  Upper Saddle River, NJ, 2001).


\bibitem{AnHeAnSi05} J.S. Andrade Jr., H.J. Herrmann, R.F.S. Andrade and L.R.da Silva,
Phys. Rev. Lett. {\bf 94}, 018702 (2005).

\bibitem{ZhCoFeRo06}
Z.Z. Zhang, F. Comellas, G. Fertin and L.L. Rong,
J. Phys. A {\bf 39}, 1811 (2006).

\bibitem{ZhYaWa05} T. Zhou, G. Yan, and B.H. Wang,
Phys. Rev. E {\bf 71}, 046141 (2005).


\bibitem{ZhZhRo06}
Z. Z. Zhang, L. L. Rong, and S. G. Zhou,
 Phys. Rev. E, {\bf 74}, 046105 (2006).


\bibitem{RaBa03}
E. Ravasz and A.-L. Barab\'asi, Phys. Rev. E {\bf 67}, 026112
(2003).

\bibitem{XiWaHuZh05}
Y.-B. Xie, B.-H. Wang, B. Hu, and T. Zhou, Phys. Rev. E {\bf 71},
046135 (2005).

\bibitem{GuMoTuAm05}
R. Guimera, S. Mossa, A. Turtschi, and L. A. N. Amaral, Proc. Natl.
Acad. Sci. U.S.A. {\bf 102}, 7794 (2005).

\bibitem{AlAlNa04}
R. Albert, I. Albert, and G. L. Nakarado, Phys. Rev. E {\bf 69},
025103(R) (2004).

\bibitem{Sc01}
G. Schliecker, J. Phys. A, {\bf 34}, 25 (2001).

\bibitem{BaRaVi01} A.-L. Barab\'asi, E. Ravasz, and T. Vicsek,
          Physica A  {\bf 299}, 559 (2001).

\bibitem{DoGoMe02} S. N. Dorogovtsev, A. V. Goltsev, and J. F. F. Mendes,
          Phys. Rev. E {\bf 65}, 066122 (2002).

\bibitem{CoFeRa04} F. Comellas, G. Fertin and A. Raspaud,
Phys. Rev. E {\bf 69}, 037104 (2004).

\bibitem{ZhRoZh07}
Z. Z. Zhang, L. L. Rong, and S. G. Zhou,
Physica A {\bf 377} (2007) 329.

\bibitem{HiBe06}
M. Hinczewski and A. N. Berker, Phys. Rev. E {\bf 73}, 066126
(2006).

\bibitem{ZhZhZo07}
Z. Z. Zhang, S. G. Zhou, and T. Zou, Eur. Phys. J. B {\bf 56}, 259
(2007) .
\end{thebibliography}
\end{document}